\documentclass[conference,a4paper]{APSIPA2025}
\usepackage{amsmath}
\usepackage{graphicx}
\usepackage{multirow}
\usepackage{threeparttable}
\usepackage{caption}
\usepackage{booktabs}
\usepackage{geometry}
\usepackage{subfigure} 
\usepackage{fancyhdr}

\fancypagestyle{firststyle}{
	\fancyhf{}
	\fancyhead[C]{2025 Asia Pacific Signal and Information Processing Association Annual Summit and Conference (APSIPA ASC)}
}

\begin{document}
	

\title{An Efficient Transfer Learning Method Based on Adapter with Local Attributes for Speech Emotion Recognition}
\author{
  \authorblockN{
    Haoyu~Song\authorrefmark{1} and
    Ian~McLoughlin\authorrefmark{1} and
    Qing~Gu\authorrefmark{2} and
    Nan~Jiang\authorrefmark{2} and
    Yan~Song\authorrefmark{2} 
  }
  \authorblockA{
    \authorrefmark{1}
    Singapore Institute of Technology, Singapore\\
    E-mail: \texttt{2303822@sit.singaporetech.edu.sg, ian.mcloughlin@singaporetech.edu.sg}\\
    \quad Tel: +65\,6592\,1189
  }
  \authorblockA{
    \authorrefmark{2}
    School of Information Science and Technology, University of Science and Technology of China, Hefei, China\\
    E-mail: \texttt{qinggu6@mail.ustc.edu.cn, jiang\_nan@mail.ustc.edu.cn, songy@ustc.edu.cn}
}
}
\maketitle
\thispagestyle{firststyle}  
\pagestyle{fancy}          

	\begin{abstract}
		Existing speech emotion recognition~(SER) methods commonly suffer from the lack of high-quality large-scale corpus, partly due to the complex, psychological nature of emotion which makes accurate labeling difficult and time consuming.
		Recently, transfer learning based methods that exploit the encoders pretrained on large-scale speech corpus (\textit{e.g.}, Wav2Vec2.0 and HuBERT) have shown strong potential for downstream SER tasks. 
		However, task-specific fine-tuning remains necessary for various conversational scenarios of different topics, speakers and languages to achieve satisfactory performance. 
		It generally requires costly encoder retraining for individual SER tasks.
		To address this issue, we propose to train an adapter with local attributes for efficient transfer learning.	
		Specifically, a weighted average pooling-Transformer~(WAP-Transformer) is proposed as a lightweight backbone to enrich the frame-level representation.
		An adapter with teacher-student branches is exploited for task-agnostic transfer learning, where the student branch is jointly optimized via mask prediction and self-distillation objectives, and the teacher branch is obtained online from the student via exponential moving average~(EMA).
		Meanwhile, local attributes are learned from the teacher branch via unsupervised clustering, which aims to act as a universal model that provides additional semantic-rich supervisions.
		A statistical attentive pooling~(SAP) module is proposed to obtain utterance representation for fine-tuning.
		To evaluate the effectiveness of the proposed adapter with local attributes, extensive experiments have been conducted on IEMOCAP.
		Superior performance has been reported, compared to the previous state-of-the-art methods in similar settings.
	\end{abstract}
	
	\section{Introduction}
	Speech Emotion Recognition~(SER) has gained considerable attention due to its potential in application scenarios like human-computer interaction, affective computing, and mental health monitoring ~\cite{schuller2011recognising,reitmaier2022opportunities}. 
	The task remains challenging, particularly in cross-session settings, where models must generalize to unseen speakers with diverse styles, prosodies, dialects, languages, and acoustic variations~\cite{LPMN2024,yin2020speaker,GSDA2023}. 

	Unlike automatic speech recognition~(ASR), where phonetic content is relatively stable across different individuals~\cite{hinton2012deep}, emotional expressions are inherently subjective and influenced by cultural background, speaker physiology, and recording conditions~\cite{ververidis2006emotional}. 
	It is thus necessary to collect a large-scale high-quality emotion corpus with wide coverage of speakers, languages and topics to guarantee the SER performance.
	However, due to the expensive annotation cost, existing SER methods~\cite{yu2020attention, ahn22b_interspeech, att3, MIL_icassp, mao19_interspeech} often struggle with few-shot learning and domain shift issues.
	
	Recently, transfer learning based SER methods, that employ pretrained Wav2Vec2.0 and HuBERT~\cite{hubert, baevski2020wav2vec} as front-end encoders, have demonstrated the ability to address data scarcity issues~\cite{w2v2pt,zhang2019multimodal,geetha2024multimodal}. 
	However, it is still necessary to retrain the encoder for individual SER tasks to achieve satisfying performance -- something computationally complex and requiring a large memory space. 
	In~\cite{zhao2024}, Ren et al. presented a self-distillation method to simultaneously fine-tune a pretrained Wav2Vec2.0 model and train a shallow version of it for fast yet effective SER.
	In~\cite{LPMN2024}, a local prototypical mapping network~(LPMN) was proposed to characterize the complex distribution of latent embedding spaces via an unsupervised learning method similar to VQ-VAE~\cite{vqvae}.
	A prototype selection scheme is used to reduce bias caused by irrelevant factors during post-processing for specific SER tasks.
	
	In this paper, we propose to learn an adapter with local attributes for various SER tasks (\textit{e.g.}, that are cross-session validated using the Interactive Emotional Dyadic Motion Capture~(IEMOCAP) corpus~\cite{IEMOCAP}). 
\begin{figure*}
	\centering
	\includegraphics[width=0.8\textwidth]{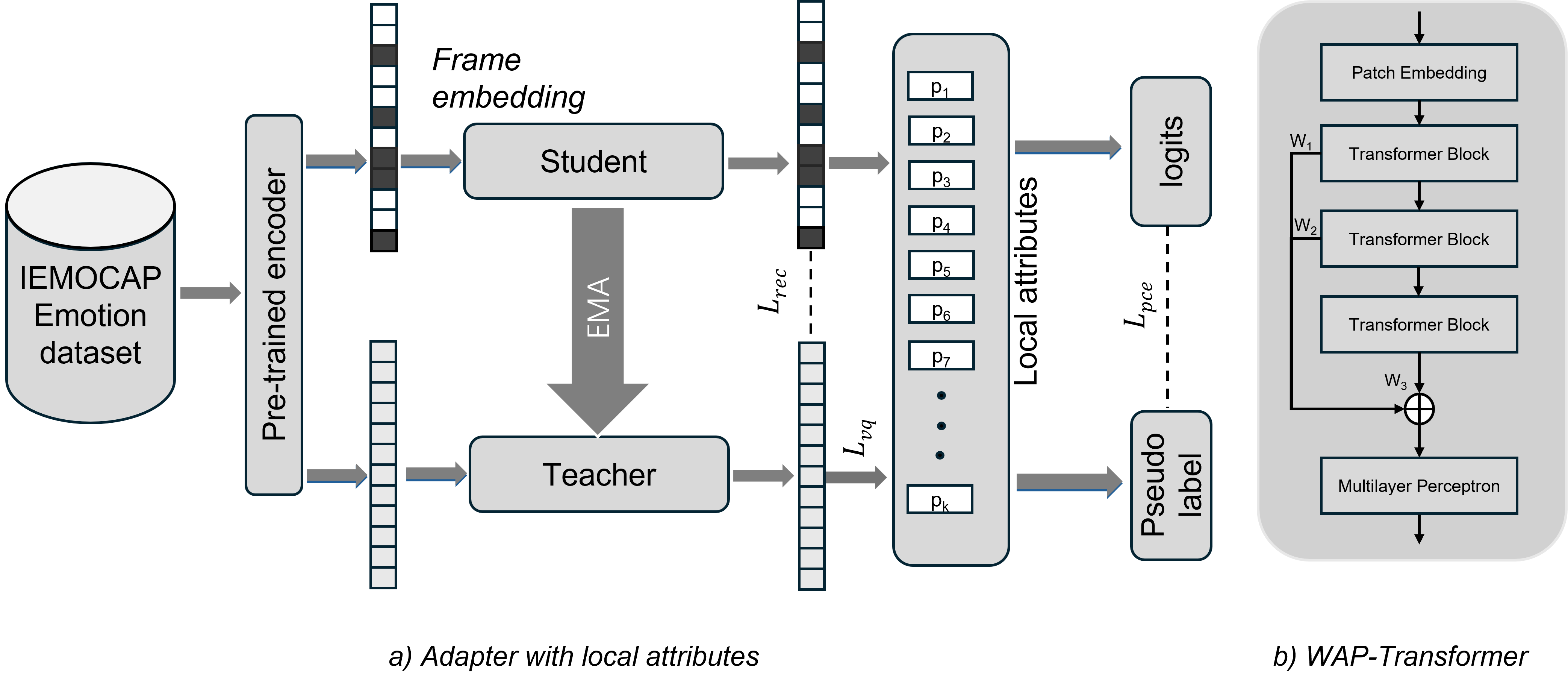}
	\caption{a) The architecture of adapter training with mask prediction and online self-distillation on extracted frame-level feature space, b) The structure of WAP-Transformer used as backbone of adapter}.
	\label{fig:overall_diagram}
\end{figure*}
	As shown in Fig.~\ref{fig:overall_diagram}, a weighted average pooling Transformer~(WAP-Transformer) is first proposed as a lightweight backbone to provide enriched frame-level representation.
	An adapter with teacher-student branches is then designed for self-supervised transfer learning.
	The student branch is optimized with a mask prediction objective similar to~\cite{he2022masked, bao2021beit,mim}, from which the teacher branch are online updated via exponential moving average~(EMA). 
        Meanwhile, we perform self-distillation on masked frames and take the teacher network as an online tokenizer,  from which the local attributes are simultaneously learned to characterize the frame-level feature distribution via batch-wise unsupervised clustering.
        In effect, the learned local attributes act as the universal model to enhance the task-agnostic adapter training for various downstream SER tasks, partly addressing the domain shift issues.
	Furthermore, a statistical attentive pooling~(SAP) is proposed to aggregate the output of the adapter into an utterance-level representation, which is classified according to the specific SER task.
	It is worth noting that only the lightweight adapter and classifier need to be finetuned, which is more efficient than retraining the while encoder.

    We perform 5-fold cross-validation with cross-session settings for evaluation~\cite{LPMN2024,xi2022frontend}.
	Experimental results on the IEMOCAP~\cite{IEMOCAP} benchmark demonstrate superior performance, achieving 78.32\%, 77.56\% and 75.63\% unweighted accuracy~(UA), weighted accuracy~(WA) and F1-score, respectively. 

	
	\section{Methodology}
	\subsection{Overview of Adapter with Local Attributes}
	The architecture of our proposed adapter with local attributes is shown in Fig.~\ref{fig:overall_diagram}, which consists of: 1) a pretrained encoder \textit{i.e.},HuBERT-large-960~\cite{hubert}, 2) a lightweight WAP-Transformer, and 3) the teacher-student branches.
		
	\subsubsection{\textbf{Pretrained encoder}}
	This consists of a 7-layer convolutional block, designed to capture low-level spectral and temporal cues, followed by a deep 24-layer Transformer-based context network with a hidden dimensions of 1024, 16 attention heads, and an intermediate feed-forward dimension of 4096. 
	Pretrained on 960 hours of speech data, this model yields robust and general acoustic representations.
	
	\subsubsection{\textbf{Lightweight WAP-Transformer}}
	This consists of 3 transformer layers, with the outputs of each layer being adaptively pooled to enrich the frame-level representation.
	The aim is to reduce the adapter model size. 
	The WAP-Transformer input is a sequence denoted by $X=X_{emb} + E_{pos}$, where
	the embeddings $X_{emb} \in R^{T \times D}$ are obtained by feeding the input waveform to the pretrained encoder. 
	Following~\cite{bao2021beit}, a standard learnable 1D position embedding $E_{pos} \in R^{T \times D}$ is added to frame-level, and $T$ is sequence length. 
	The output of the $l$-th transformer layer is conceptually defined as $X^l = Transformer(X^{l-1})$.
	The output of WAP-transformer $Z$ is weighted averaged $Z=\sum_{l=1}^L w_l X^l$ with learnable parameters $W=[w_1,..., w_L]$, where $L=3$,
	 
	\subsubsection{\textbf{Teacher-student branches}}
	A Siamese structure with lightweight WAP-Transformer as backbone is exploited to perform self-supervised transfer learning on a given emotion corpus.
	As aforementioned, the student branch is optimized with both mask prediction and self-distillation objectives, and the teacher is ensembled from the students over epochs. 
	The training process is detailed as follows, consisting of mask prediction, local attribute learning and frame-level self distillation.
	
	\subsection{Adapter training}
	\subsubsection{\textbf{Mask Prediction}}
	The input to the student branch $X'$ is the masked version of sequence $X$, where 40\% of positions $M$ are randomly selected and replaced with learnable embedding $e_{[M]} \in R^D$.
	By feeding $X'$ to the student branch \textit{i.e.}, WAP-Transformer $E_{\theta}^s$, the reconstructed sequence $Z^s=\{z_i^{s}\}_{i=1}^T$ is obtained by predicting from the remaining visible ones according to context information.
	The input to the teacher branch $E_{\theta}^t$ is $X$, and the sequence $Z^t=\{z_i^{t}\}_{i=1}^T$ is used as the target of the masked prediction for optimizing the student parameters $\theta^s$.
	The mask reconstruction loss is formulated as the mean square error of masked positions, which is
	\begin{equation} \label{eq:rec}
		\mathcal{L}_{\text{rec}} = \frac{1}{\|M\|} \sum_{i \in M} \|z_i^{t} - z_i^{s} \|_2^2
	\end{equation}
	And the parameter of teacher $\theta^s$ is updated using EMA, that is
	\begin{equation}
		\theta_t^{i+1} = \alpha\theta_t^i + (1-\alpha)\theta_s^{i+1}
	\end{equation}
	where $i$ denotes the learning iteration, and $\alpha$ is the smoothing hyperparameter, set to 0.999 by default.
    The large $\alpha$  is used to smooth the rapidly changing student parameters. This makes the teacher model more stable, avoids overfitting to noise, and provides a consistent signal that improves convergence and generalization.

	This EMA smooths out the short-term noise in the student’s rapidly changing parameters, producing a more stable teacher signal that accelerates convergence and improves generalization.

	In~\cite{bao2021beit,hubert}, an offline additional tokenizer is further introduced to represent the utterance as a sequence of discrete tokens.
    We propose to learn the local-attribute prototypes directly from the teacher branch; these prototypes offer semantically rich supervision that complements the masked-prediction objective.
    Because each prototype is the quantiser output, its index
    $k(\mathbf{z}^{t})$ naturally serves as the pseudo-label for frame-level self-distillation, while the student produces its logits by computing the (negative) distances between its own embedding and all prototypes (see next subsection).
			
	\subsubsection{\textbf{Local Attributes Learning}}
	Local attributes $P = \{p_1, p_2, \ldots, p_K\}$ are learned to characterize the frame-level feature distribution, which can be learned by a batch-wise unsupervised clustering method.
	Given a batch of the extracted frame-level embeddings $Z^t$ from the teacher branch, local attributes $P$ can be learned as follows
	\begin{equation} \label{batch}
        \mathcal{L}_{\text{VQ}} = \sum_{k=1}^{K} \sum_{z^t \in p_k} \| z^t - p_k\|_2^2
	\end{equation}
	where $z^t \in p_i$ denotes embeddings nearest to the $i-th$ attribute $p_i$. 
	That is, we can assign $z^t$ to pseudo-label $i$, according to Euclidean distance 
	\begin{equation} \label{eq:pseudo}
		i = \arg\min_k\|z^t - p_k\|_2^2, 
	\end{equation}
	And the corresponding attribute is updated as 
	\begin{equation} \label{update}
		p'_i \leftarrow p_i + \eta_t(z^t - p_i)
	\end{equation}
	The learned attributes may be considered as a universal model to characterize the distribution of frame-level embeddings.
	We further perform self-distillation to update the student branch with local attributes, besides the reconstruction loss $L_{rec}$ in eqn.\eqref{eq:rec}.
	
        \subsubsection{\textbf{Frame-level Self-Distillation}}
	We perform frame-level self-distillation on the masked positions $M$.
	From the teacher branch, we can obtain the pseudo-labels of $Z^t$ according to eqn.~\eqref{eq:pseudo}.
	For the student branch, the logits can be obtained by computing the cosine similarity between of embeddings $Z^s$ and local attributes $P$. 
	 The pseudo cross entropy~(CE) loss $\mathcal{L}_{pce}$ can then be computed to jointly optimize with $\mathcal{L}_rec$, that is
	\begin{equation} \label{eq:losst}
		\mathcal{L} = \lambda\mathcal{L}_{rec} + (1-\lambda)\mathcal{L}_{pce}
	\end{equation}
	where $\lambda$ is the hyperparameter for balance reconstruction and pseudo CE loss. 
    A ramp-down scheme from $1.0$ to $0.5$ is applied to place more emphasis on $\mathcal{L}_{rec}$ at the early stage, allowing the student to first learn stable representations by mimicking the teacher. As training progresses, the weight gradually shifts toward $\mathcal{L}_{pce}$, enabling the model to focus more on discriminative learning. The lower bound of $0.5$ ensures that $\mathcal{L}_{rec}$ still contributes throughout training, preventing the student from completely drifting away from the teacher guidance.

    

        \subsection{Finetune on SER task}
	We follow~\cite{xi2022frontend,LPMN2024} to further finetune the adapter for SER.
	An SAP module comprising a learnable convolution and a statistic pooling layer is proposed to exploit frame-level attention.
	The convolution layer weighs the frame-level features, while a statistic pooling layer collects the $1^{st}$- and $2^{nd}$-order weighted statistics.
	Specifically, let $Z \in R^{T \times D}$ be the output features for $L$ frames with $D$ dimensions.
	By feeding $Z$ into a $1 \times 1$ convolution layer, followed by softmax activation along the time axes, an attentive weight map $A \in R^T \times K$ can be obtained, with $K$ attention heads.
	Statistical pooling is performed:
	\begin{equation} \label{eq:stat_pooling}
		\begin{split}
			\boldsymbol{\mu} &= Z^TA \\
			\boldsymbol{\sigma^2} &= \textbf{Z}^T(\textbf{A}\odot \textbf{A}) - (\textbf{Z}^T\textbf{A} \odot  \textbf{Z}^T\textbf{A})
		\end{split}
	\end{equation}    
	where $\odot$ denotes the Hadamard product. The $\boldsymbol{\mu}$ and $\boldsymbol{\sigma^2}$ are $L_2$-normalized and concatenated as the input the final emotion classification layer.
	IEMOCAP 5-fold cross-validation is used for fine-tuning, where the dataset is divided into five sessions (each with 2 speakers, for a total of 10 speakers). 
	In each fold, one session serves as the validation set while the remaining four sessions are used for training.

\begin{figure*}[!tbp]
      \centering
      \subfigure[]{
          \includegraphics[width=0.4\linewidth]{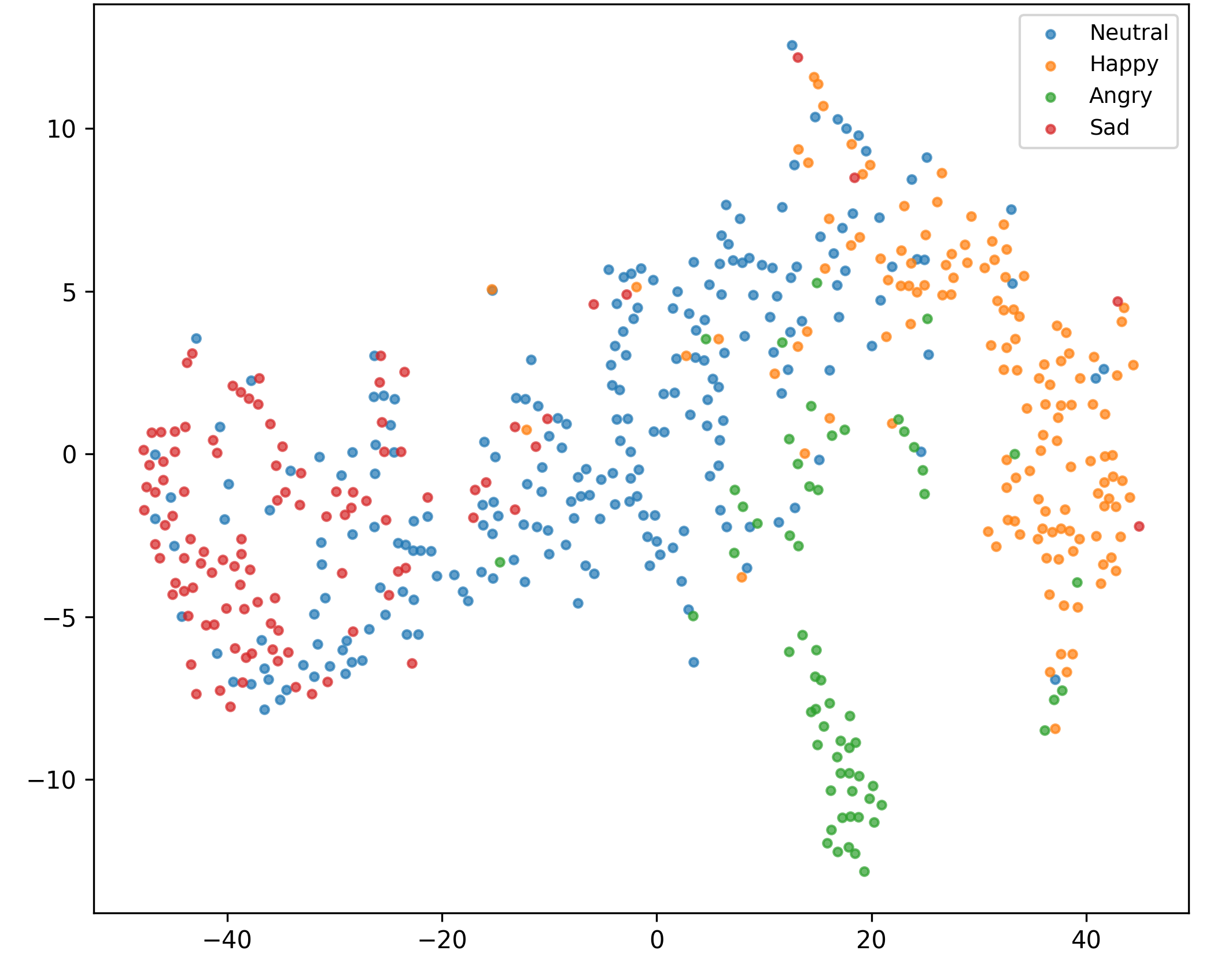}
          \label{fig:tsne_without}
      }
      \hfill  
      \subfigure[]{
          \includegraphics[width=0.4\linewidth]{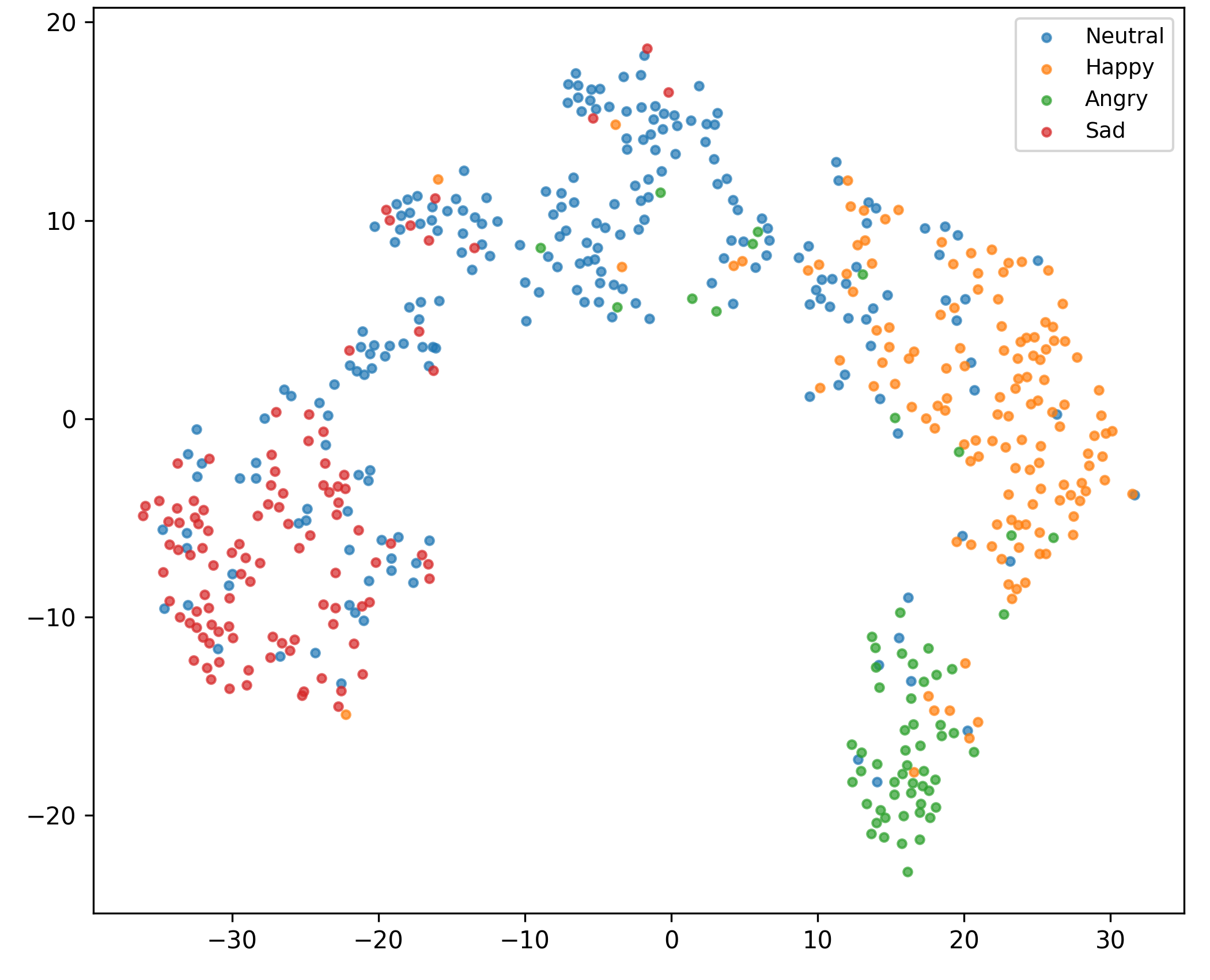}
          \label{fig:tsne_with}
      }
      \caption{t-SNE projections emotion embeddings on unseen test data.
      (a) Emotion embeddings obtained from pretrained encoder,  
      (b) Emotion embeddings obtained from our proposed adapter.}
      \label{fig:combined_embeddings}
\end{figure*}
	

    \section{Experiments}
	\subsection{Dataset and Evaluation Metrics}
	IEMOCAP contains recordings of multiple actors in improvised and scripted dialogues with a wide variation in speaker and session characteristics~\cite{IEMOCAP}. 
	It is widely used for SER research and allows us to compare the effectiveness of adapter to other systems. 
    We assessed performance using WA, UA, and macro F1 scores.

	WA computes a weighted average of per-class accuracy, while UA, as a simple average, better reflects fairness in the presence of class imbalance. 
The macro F1-score, computed as the average F1 across all classes, provides a balanced evaluation by giving equal weight to each class, which is particularly useful under class imbalance.  

    The results, presented below, confirm the effectiveness of proposed data augmentation strategy in enhancing robustness.
	
	\subsection{Implementation details}
	As aforementioned, the adapter module exploits the WAP-Transformer backbone, acting on the output of the pretrained encoder (HuBERT-large-960), as shown in Fig.~\ref{fig:overall_diagram}. 
	This begins with a patch embedding layer that projects 1024-dimensional inputs to 384 dimensions, followed by three Transformer blocks at the 384-dimensional level. 
	A fourth block is a 3-layer MLP  then applied for aggregation.
	A Siamese network based WAP-Transformer is designed as adapter, and corresponding $K$ local attributes are learned as a universal model for self-distillation.
	A SAP module with 4 attention heads is used to project a sequence of frame-level embeddings into utterance-level representation for SER.
	
	Adapter training is conducted with a batch size of 96 over 100 epochs using the Adam optimizer with an initial learning rate of 1e-4 and a cosine learning rate scheduler. 
	Fine-tuning also employs a batch size of 96 over 100 epochs, but uses cross-entropy loss for the primary emotion classification task with improvised dialogues, over five classes (with happy and excited merged to account for limited data). 
	To further address data imbalance and leverage the reconstruction ability of our trained model, we apply data augmentation during fine-tuning by randomly masking parts of the embeddings for minority classes, thereby enhancing sample diversity and balancing class distributions. 
	
\subsection{Comparison with state-of-the-art methods}
	Table~\ref{tab:iemocap_results} presents 5-fold cross validation results in terms of UA, WA and F1-score, for adapter training and several state-of-the-art systems.
	Overall results demonstrate that our adapter can outperform existing methods on the challenging IEMOCAP dataset. 
	We achieved a UA of 78.32\% and an WA of 77.56\%, surpassing previous approaches such as Co-attention, W2v2-PT~\cite{coattn}, Spk-norm~\cite{speakernorm}, GLRF~\cite{ding2023GLRF}, and LPMN~\cite{LPMN2024} in similar settings. 
	For example, The competitive LPMN reported UA and WA values of 76.85\% and 74.50\% respectively.
	After a prototype selection for post processing, it can improve to 77.42\% and 75.82\%. 
	Our proposed WAP-Transformer improves over LPMN-1024-s60 around 0.80\% and 2.26\% in terms of UA and WA metrics.
	These gains are particularly impressive given the inherent challenges of IEMOCAP, which include significant cross-session and cross-speaker variability that typically degrade performance.
	The results validate the effectiveness of combining a large-scaled pre-trained encoder, WAP-Transformer based adapter training, and fine-tuning.
\begin{table}
    \centering
    \scriptsize   
    \caption{Validation scores of the proposed WAP-Transformer compared with state-of-the-art methods on IEMOCAP.}
    \begin{tabular}{lccc}
        \toprule
        \textbf{Method} & \textbf{UA} & \textbf{WA} & \textbf{F1} \\
        \midrule
        Co-att~\cite{coattn}          & 72.70 & 71.64 & --    \\
        W2v2-PT~\cite{w2v2pt}         & --    & 67.20 & --    \\
        Spk-norm~\cite{speakernorm}   & --    & 74.20 & --    \\
        GLRF~\cite{ding2023GLRF}      & 73.31 & 72.81 & 72.92 \\
        LPMN-1024-s60~\cite{LPMN2024} & 77.42 & 75.82 & 75.71 \\
        LPMN~\cite{LPMN2024}          & 76.85 & 74.50 & 74.88 \\
        WAP-Transformer (Ours)        & 78.32 & 77.56 & 75.63 \\
        \bottomrule
    \end{tabular}
    \label{tab:iemocap_results}
\end{table}

    Fig.~\ref{fig:combined_embeddings} compares the utterance-level embedding spaces obtained from the same HuBERT encoder before and after frame-level self-distillation.
    (a) Without self-distillation, the clusters corresponding to the four emotions are loosely formed and partially overlapped -- particularly between Neutral and Happy.
    (b) Introducing frame-level self-distillation leads to markedly tighter intra-class cohesion and larger inter-class margins, yielding a much cleaner separation of Sad and Angry from the other emotions while also reducing the number of outliers.
    These observations confirm that self-distillation effectively drives the encoder to focus on affect-relevant acoustic cues, thereby enhancing the discriminability of the learned representations for downstream SER.
	\subsection{Ablation study to explore key components}
	To evaluate the impact of the key contributions in WAP-Transformer, we conduct an ablation study focusing on three critical factors:  
	(1) the effect of adapter training via WAP-Transformer,  
	(2) the influence of local attribute (LA) size in the VQ module
	Table~\ref{tab:ablation_summary} summarizes the performance under different configurations.

Overall, these ablation experiments reveal some useful findings. 
Firstly, adapter training can significantly boost performance with WAP-Transformer. 
Without adapter training, the model achieved a UA of 75.73\%, a WA of 74.91\%, and an F1 score of 72.90\%.
All metrics increased after adapter training, to 78.32\% (UA), 77.56\% (WA), and 75.63\% (F1) respectively -- around 2\% absolute increase for each. 
This confirms that adapter training benefits from WAP-Transformer to learn local attributes, capturing semantic-rich information like speakers, phonetic content and styles. 

\begin{table}[t]                   
	\centering
	
	\caption{Ablation studies on the impact of WAP\textendash LA components and training settings.}
	\label{tab:ablation_summary}       
	
	\begin{tabular}{p{3.5cm}ccc}
		\toprule
		\textbf{$K$ attributes} & \textbf{UA (\%)} & \textbf{WA (\%)} & \textbf{F1-score} \\
        \midrule
		$K\!=\!1024$  & 78.32 & 77.56 & 75.63  \\
		$K\!=\!512$  & 77.22 & 76.93 & 74.26 \\
		$K\!=\!256$  & 77.00 & 76.21 & 75.14 \\
		$K\!=\!128$  & 76.10 & 76.28 & 73.80 \\
		
		\midrule

		WAP-Transformer \\(w/o Adapter Training) & 75.73 & 74.91 & 72.90 \\

		\bottomrule
	\end{tabular}
	
	\vspace*{-3pt}                     
\end{table}
When we consider local attributes~(LA), Table~\ref{tab:ablation_summary} indicates that a size of 1024 yields the best performance of 78.32\% and 77.56\% and 75.63\% in UA, WA and F1-score. 
Reducing the LA size degrades performance, suggesting that a larger size is more effective at capturing fine-grained details, although it may also increase computational cost and the risk of overfitting.

Ablation studies collectively confirm that the component of the proposed approach -- adapter training, an appropriate LA size, and careful integration of domain information all play a crucial role in improving performance. 

\section{Conclusion}
This paper has proposed WAP-Transformer to effectively reduce the gap between a large-scale pretrained HuBERT model and downstream SER tasks with limited corpus size. WAP-Transformer introduces a lightweight transformer (\textit{i.e.}, WAP-Transformer) based adapter, which contains a teacher and student branch.
WAP-Transformer learns local attributes that can effectively capture the semantic-rich information related to speakers, content and styles for different tasks.
After adapter training and adjustment, we finetune the adapter to perform SER tasks via an attentive pooling layer and a classification layer.
5-fold cross validation on IEMOCAP demonstrates superior performance of 78.32\%, 77.56\% and 75.63\% for UA, WA and F1-scores respectively. 
Ablation studies confirm the effectiveness of the main contributions introduced by WAP-Transformer. 

\bibliographystyle{ieeetr}
\bibliography{mybib}

\end{document}